# Toward a Deeper Understanding of General Relativity

John E Heighway
NASA (retired)
1099 Camelot Circle
Naples, Florida 34119
jeheighway@hotmail.com

**Abstract**

Standard treatments of general relativity accept the gravitational slowing of clocks as a primary phenomenon, requiring no further analysis as to cause. Rejecting this attitude, I argue that one or more of the fundamental "constants" governing the quantum mechanics of atoms must depend upon position in a gravitational field. A simple relationship governing the possible dependencies of e, h, c and $m_e$ is deduced, and arguments in favor of the choice of rest mass, $m_e$, are presented. The reduction of rest mass is thus taken to be the sole cause of clock slowing. Importantly, rest mass reduction implies another effect, heretofore unsuspected, namely, the gravitational elongation of measuring rods. An alternate ("telemetric") system of measurement that is unaffected by the gravitational field is introduced, leading to a metric that is conformally related to the usual proper metric. In terms of the new system, many otherwise puzzling phenomena may be simply understood. In particular, the geometry of the Schwarzschild space as described by the telemetric system differs profoundly from that described by proper measurements, leading to a very different understanding of the structure of black holes. The theory is extended to cosmology, leading to a remarkable alternate view of the structure and history of the universe.

## 1 Introduction

To date, the general theory of relativity has been understood in terms of a single interpretation: It tells us what a local observer will measure using local clocks and measuring rods. Such information must of course be provided by any coherent theory: it is absolutely necessary. At first thought, it is difficult to imagine any alternative system for making measurements. However, certain phenomena strongly suggest that the characteristics of local measuring devices are not constant, but rather depend upon position in a gravitational field. In particular, it is a well-established fact that the rate of a clock is reduced by the presence of nearby massive objects. Einstein predicted this effect in his 1916 paper.[1] Elsewhere he speculated regarding possible changes in the length of measuring rods,[2] and, regarding the possible effect of gravity on the masses of objects, concluded that the inertial mass of a test object would be increased when ponderable masses approach.[3] Finally, one often sees it suggested that the speed of light is reduced in a gravitational field.[4] Thus, the delay of radar echoes from planets near superior conjunction with the Sun (the Shapiro effect) is sometimes explained in terms of this purported reduction in the speed of light.[5]

Here it is logical to ask the question, "What system of measurement is being tacitly appealed to in such discussions regarding the rates of clocks, the lengths of measuring rods, the masses of objects, and the speed of light?" It clearly cannot correspond to the measurements made by local observers, for, in the first three instances, we are considering changes to the very standards by which measurements are made. In the case of the speed of light, it is an axiom of the theory that the speed of light will be invariant as measured by local observers.

What is needed is a system that is not influenced by gravitational fields, so as to permit the inter-comparison of two objects, which are identical when brought together, in situations where they are remote from one another. But is such a system of remote inter-comparison possible? At present, the preponderance of opinion seems to be in the negative. Thus, Robert Dicke has remarked,[6]

> (The statement that) a hydrogen atom on Sirius has the same diameter as one on Earth … is either a definition or else is meaningless.

This pessimism is unjustified, for in certain situations, such an inter-comparison is possible. In fact, whenever the gravitational field does not change with time, the rates of clocks that are remote from one another can be compared. This is the case with the famous Schwarzschild solution to Einstein's field equations describing the gravitational field outside an isolated, spherically symmetric, non-rotating, massive body.

## 2  Gravitational Clock Slowing

Consider two observers situated along a single radial line in such a field. If a continuous light signal produced by a certain atomic transition is sent from the lower observer to the higher observer, a red shift of frequency will be observed by the latter.

Here we encounter a little heresy that has somehow crept into physics. It is sometimes incorrectly claimed that the red shift may be understood to be the result of a change in the frequency of the photons that constitute the signal as they "climb up against the pull of gravity." The fact is that the photons were emitted at a reduced frequency – even lower than that measured by the higher observer – because of the gravitational slowing of clocks. The wavecrests of an electromagnetic wave are conserved: the wave history at any point along the trajectory of the wave is a faithful repetition of the emitted wave history. No wavecrests are created en route, nor are any lost. One may think of them as being analogous to the mantras repeated by a holy man. If they are received at a slower pace than is usual, it is because they were emitted (or recited) at a slower pace than is usual. Wavecrest or mantras, there must be, for a continuous signal in a time-independent field, a fixed number of them between source and observer: For each one that enters the space separating source and observer, there must be one leaving that space. Otherwise, the field would not be constant in time. The true frequency must be the same, and any difference in the measured frequency must be attributed to the fact that observers are measuring frequencies using "standards" that are reduced by the action of the gravitational field.

For those who remain unconvinced by this simple yet elegant argument, originally given by Einstein in a 1911 paper,[7] it may be said that this question has already been decided experimentally, in effect, by the wonderful technology of the Global Positioning Satellite System.

Also, although to date no one seems to have carried out a simple, direct experiment to settle the issue, it is clear that Cesium clocks now possess the required accuracy to prove directly that the red shift is fully accounted for by the slowing of clocks, and hence, that free photons proceed without any change in frequency as they move in a constant gravitational field.

## 3 A Telemetric System of Measurement

This constancy of the frequency of freely moving electromagnetic waves is the key to the implementation of a system of space-time measurement that is unaffected by the gravitational field. One simply measures time using the signals from a single remote clock. Distance measurements are then made (by local or remotely located observers) using electromagnetic echo ranging (radar) techniques, calculated using the time as measured, not by a local clock, but by the same remote clock.

Since the new system of measurement employs a remote clock and makes use of a remote sensing technique for distance measurements, the name *telemetric* is suggested. As will become clear, the insights provided by this scheme will naturally result in an extended interpretation of the general theory, which, for reasons that will become clear, will be called the Variable Rest Mass (VRM) Interpretation.

## 4 Why Are Clocks Slowed?

In specifying how distance is to be measured, we have jumped ahead in our reasoning, since this technique assumes that the speed of light may be taken to be a true constant. In order to justify this assumption, it is necessary that the cause of the slowing of clocks be systematically investigated. The frequency of an atomic transition depends upon the four fundamental constants of quantum electrodynamics: the electronic charge, e; the speed of light, c; Planck's constant, h; and finally, the mass of the electron, $m_e$. (The masses of all fundamental particles are assumed to be influenced, if at all, in exactly the same manner.)

In seeking the cause of the gravitational slowing of clocks, it is assumed that one or more of these parameters, which are indeed constants when measured in the usual way, must, in the new system, be dependent upon position in a gravitational field. Quantities measured in the new telemetric system will be indicated by appending an asterisk. Quantities measured in the usual way by local observers using local clocks and measuring rods are called proper measurements, and are written plain. Quantities that are measured to be the same in both systems will also be written plain.

In addition to accounting for the slowing of clocks, the variability of these parameters is constrained by two facts: first, the conservation of electric charge – which fixes e*, and so removes it from consideration; second, that all local observers measure the apparent speed of light to be constant. It turns out that there are just two reasonable ways to satisfy the above requirements, as detailed in Appendix A. The simplest requires that the rest mass of the electron, $m_e$* (and that of all particles), is reduced in a gravitational field. For the Schwarzschild field the factor is

$$f = \sqrt{(1 - 2GM/c^2 r)}.$$

Here M is the mass of the source of the gravitational field, G is the Newtonian constant of gravitation, r is a radial coordinate, and c is the speed of light. One alternative is that the speed of light, c*, is reduced by the factor f, while Planck's constant, h*, increases by the reciprocal factor, $f^{-1}$, $m_e$* increases by $f^{-2}$, and atomic dimensions, and hence the length of measuring rods, are unaffected.

As previously mentioned, the decrease in the speed of light has been widely accepted. Indeed, Einstein, having correctly understood the gravitational red shift in terms of the slowing of clocks, seems to have evoked the reduction of the true speed of light in order to explain why local observers, in spite of their slowed clocks, will always to get the same answer when measuring the speed of light regardless of their position in the field. Here he evidently made the unconscious assumption that the length of measuring rods would be unaffected.

But the detailed analysis of Appendix A shows that if $c^*$ is decreased, then $h^*$ must be increased by the reciprocal factor. If it is non-intuitive that the speed of light should be affected by gravity, it is downright mysterious that Planck's constant should also be affected. In contrast, the dependence of the rest mass of objects upon their position in a gravitational field is a perfect expression of the reality of the concept of gravitational potential energy in what is usually called Newtonian physics.

If the proper rest mass of a body is measured by a local observer to be m, then its rest mass in the telemetric system will be designated by $m^*$, and we have for the rest energy in the new system

$$E^* = m^* c^2 = f\, mc^2 =$$

$$\sqrt{(1-2GM/c^2 r)}\, mc^2 \approx mc^2 - GmM/r,$$

where the final approximation gives the usual Newtonian expression. Note that the asterisk has been omitted on the symbol, c, since the speed of light is taken to be unaffected. The plain m is the proper rest mass as measured by local observers. It is a constant, and is equal to $m^*$ when the mass is infinitely removed from the large mass, M, that produces the field.

## 5   Gravitational Rest Mass Reduction

In view of the above happy correspondence with Newtonian physics, and for reasons that will be presented directly, the choice of rest mass as the "culprit" is taken, and gravitational rest mass reduction is identified as *the* fundamental phenomenon in the VRM interpretation of the general theory of relativity. It is identified as the cause, not only of the slowing of clocks, for which the comparison of rates of clocks remote from one another is directly observable, but also of another phenomenon, gravitational length dilation, which is not.

## 6   Gravitational Length Dilation

Fundamental quantum lengths – typified by the Bohr radius – depend inversely upon the masses of elementary particles. Since rest masses are reduced by the factor, f, in a gravitational field, the Bohr radius, and hence the length of all objects, must increase by the factor 1/f.

That length dilation must occur may also be understood without any quantum considerations. If clock rates are reduced, while the apparent speed of light, as well as the true speed of light is unaffected, it is clear that the length of the path over which the light travels must be underestimated; hence measuring rods must be elongated by the action of the gravitational field.

Note that all speeds, not just the speed of light, will be measured to be the same regardless of the measuring technique, whether using local clocks and rods, or using the telemetric scheme.

## 7 Energy Conservation

Corroborating the rest mass reduction interpretation is an exact integral for the motion of a test mass in the Schwarzschild field.[8] (In fact, this integral is valid in any field that is constant in time, most notably, for the Kerr solution, which describes a rotating black hole.) For the Schwarzschild field, this integral may be written

$$f/\sqrt{(1-v^2/c^2)} = \text{constant}.$$

Multiplying each member by the constant $mc^2$, using $m^* = fm$, and recalling that v and c are the same in either system of measurement, we have

$$m^*c^2/\sqrt{(1-v^2/c^2)} = \text{constant},$$

which states that the energy of the test mass, as assessed according to the VRM interpretation, is constant. It may be said that in a gravitational field, rest-mass energy is converted to kinetic energy, and vice-versa. Again, we recover a reassuring agreement with Newtonian ideas and common sense.

## 8 The Geometry of a Black Hole: Two Worlds Connected by a Flat Sphere

We now turn our attention to the geometry of a simple, non-rotating black hole, which is also described by the Schwarzschild solution.
The gravitational length dilation effect profoundly alters the perceived geometry of a black hole. The proper invariant four-interval for the Schwarzschild solution may be written in terms of the usual coordinates as[9]

$$ds^2 = f^2c^2dt^2 - f^{-2}dr^2 - r^2(d\theta^2 + \sin^2\theta d\phi^2).$$

Here r is defined as the proper circumference (i.e., that measured by local observers using local measuring rods) divided by $2\pi$, while $\theta$ and $\phi$ are the usual angles in spherical coordinates. Since in the telemetric system, both time and distance measurements are increased by the factor, $f^{-1}$, relative to corresponding proper measurements, the telemetric invariant interval is

$$ds^{*2} = f^{-2} ds^2$$

$$= c^2dt^2 - f^{-4}dr^2 - f^{-2}r^2(d\theta^2 + \sin^2\theta d\phi^2).$$

Note that the world time, t, is the same as the telemetric time. Because of the length dilation effect, the circumference of a circle will be larger when measured in the telemetric system by the factor, $f^{-1}$. Evidently then, the area of a sphere will be, in telemetric measure, equal to

$$4\pi r^{*2} = 4\pi r^2 f^{-2} = 4\pi r^2/(1-r_s/r),$$

where $r_s = 2GM/c^2$ is the Schwarzschild radius. Differentiating with respect to r, one has

$$d/dr\,[r^2/(1-r_s/r)] = (2r-3r_s)/(1-r_s/r)^2.$$

Thus the area of a sphere is not a monotone function of r; it has a minimum at $r = 3/2\, r_s$. Furthermore, inside this sphere of minimum area, the area of a sphere increases without limit as r approaches $r_s$. Also, the telemetric distance in the radial direction between $r_1$ and $r_2$ ($r_1 < r_2$) is the integral from $r_1$ to $r_2$ of $f^{-2}(r)$, which is easily calculated to be

$$R^*(r_1, r_2) = r_1 - r_2 + r_s \ln[(r_2 - r_s)/(r_1 - r_s)],$$

and this also increases without limit as $r_1 \to r_s$. Thus from the VRM point of view, the Schwarzschild solution is seen to comprise two infinite three spaces which are joined by a wormhole-like structure. At the "narrows", where $r = 3/2\, r_s$, we find a flat (as we shall see) yet spherical surface, which may be called the stenosphere (Greek, στενος = narrow). The outer world, our universe, lies outside the stenosphere, and is asymptotically Euclidean, whereas inside the stenosphere ($3/2\, r_s > r > r_s$), in what may be called "innerspace," the character of space is hyperbolic, since the measure of circumference, $r^* = f^{-1}r$, diverges more rapidly than the radial distance measure, $R^*$.

In the VRM picture, a collapsing star does not suffer an infinite compression, producing a singularity hidden behind an event horizon; it produces a sort of rupture in space, forming a connection to a distinct infinite realm, "innerspace," as it has been called, into which it falls. The event horizon, where the coordinate r has the value $r_s$ is, according to the VRM interpretation, infinitely distant, and the region within (beyond) it has exactly the same insignificance as the region "beyond infinity" has in our universe.

It would seem that these observations are in conflict with the admitted fact that an observer, freely falling along a radial line, will reach and penetrate the event horizon in a finite period of proper time, as measured with his own co-moving clock. This in spite of the fact that all fixed observers will agree that he will never in all eternity reach the horizon. In order to resolve this issue, recall the exact integral of motion for a body moving in the Schwarzschild field. It may be written

$$\sqrt{(1- r_s/r)}\, /\sqrt{(1-v^2/c^2)} = \text{constant}.$$

It is clear that the velocity of the infalling observer will go to the speed of light, $v \to c$, as $r \to r_s$. Furthermore, in the VRM interpretation, as his coordinate, r, approaches $r_s$, the observer's rest mass will go to zero, as will the rate of his clock.

Thus the region inside the event horizon may be judged to be a part of our universe only if we are willing to accept the idea that the time measured by a clock which has ceased to run has meaning, and that an observer who has ceased to exist as material being (and who never, in the infinity of our time, arrives at his destination) is a valid witness.

## 9  Possible Connections

An intriguing question remains regarding black holes; namely, whether two or more black holes share a common "innerspace," or whether each such is a separate, infinite three space. If one "innerspace" serves for all, then the possibility emerges, however impractical it may be, of travel from one black hole to another via an "innerspace" route.

## 10  Proper and Telemetric Measurements

At this point, it seems appropriate to point out an important difference between the telemetric and the proper schemes of measurement. The general theory is often described as the theory of spacetime, and the proper invariant interval refers only to spacetime measurements. It is, then, a remarkable fact that from this invariant interval one can deduce the dynamical behavior of massive bodies. One may ask how it is possible that geometry-chronometry can contain dynamics. From the VRM point of view, the answer may be offered that the dynamics enters via the dependence of rods and clocks, the instruments used to make spacetime measurements, upon the gravitational potentials.

In contrast, the telemetric invariant interval is pure spacetime. Thus, for instance, the geodesic equations derived from the telemetric metric will not reveal any gravitational force (this follows immediately from the fact that $g_{00}$ is a constant). This is just what one would expect, since the telemetric system was constructed to mimic an imaginary system of measurement employing rods and clocks that are not influenced by the gravitational field.

The VRM extended interpretation embraces both systems of measurement. On the one hand, from the proper metric, one obtains all of the relevant dynamics; on the other, the telemetric measurement scheme provides the basis for a deeper understanding of the physics, and the associated metric yields what may be said to be the "true" geometry of space.

## 11 The Embedding Diagram

In order to clarify the nature of the geometry of the Schwarzschild field, it will be helpful to introduce what is called an embedding diagram. In this diagram, the section $\theta = \pi/2$ will be shown (all central sections are identical). In the absence of the central mass that produces the field, such a section would be a simple Euclidean plane, but with the mass present, the surface must be permitted to bulge up or down – down is the usual choice – so as to correctly depict the relation between small distances, $dR^*$ and $r^*d\phi$, in the radial and circumferential directions, respectively. Because of the spherical symmetry, the surface is a figure of revolution of which only the right half of a vertical section will be shown in fig.1.

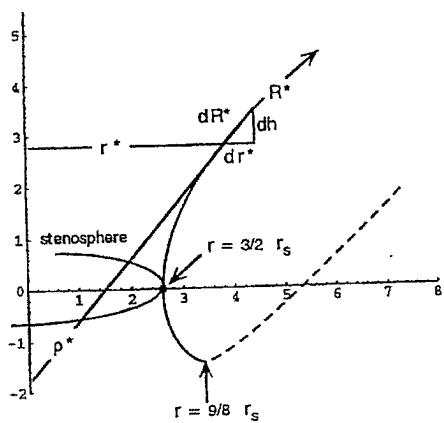
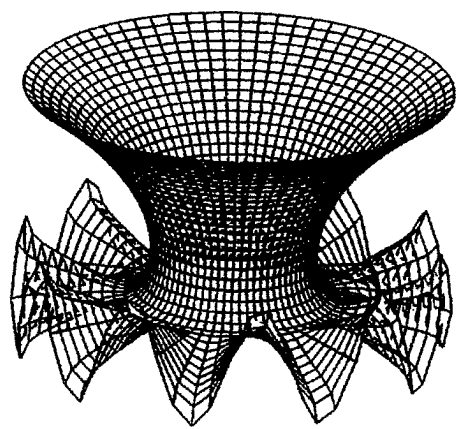

Fig. 1
Telemetric embedding diagram for the Schwarzschild field: right half of cross-section.

Fig. 2
Telemetric embedding diagram for the Schwarzschild field.

In this diagram, $r^*$ ($1/2\pi$ times the telemetric circumference of circles) is the abscissa; the parameter, h, which has no direct meaning, is the ordinate; $R^*$ is the telemetric distance along the surface in the radial direction; and the coordinate, r, now serves merely as a convenient parameter.

It is important to understand that points off the diagram surface are not part of the three space of the Schwarzschild field, and that, regarding the curvature evident in the diagram, only that component lying in the tangent cone is sensible to observers in the three space.

The differential equation for the surface in the embedding diagram is developed from the relation $dh^2 = dR^{*2} - dr^{*2}$, as indicated in figure 1, along with the relations $r^* = f^{-1} r$ and $dR^* = f^{-2} dr$. The result expressed in terms of the parameter r is

$$dh/dr = f^{-3} \sqrt{[2r_s/r \, (1 - 9/8 \, r_s/r)]}$$

It will be noted that when $r < 9/8 \, r_s$, the expression for dh/dr becomes imaginary. This is a consequence of the fact that in the infinite region ($9/8 \, r_s > r > r_s$), $r^*$ increases faster than $R^*$. In this region, the embedding diagram should actually have the appearance of a floppy hat, with a circumferential excess of material, as shown in figure 2. In figure 1, for the sake of clarity, the imaginary values for h are accepted, and are indicated by the dashed part of the curve in the diagram.

The sensible radius of curvature, $\rho^*$, is the distance from the point in question along the tangent of the surface to the central axis. On the stenosphere, $\rho^*$ becomes infinite, indicating that this surface is indeed flat. This is consistent with the well-known fact that (unstable) photon orbits occur at this location, $r = 3/2 \, r_s$.

Note also that a circle when viewed from what is conventionally denoted as the "outside" will be judged convex if it is outside the stenosphere, but will appear to be concave if it is inside the stenosphere.

## 12 The Abramowicz Effect

This makes clear the reason underlying the startling paradox regarding the direction of the "centrifugal" force experienced by bodies as they move in circles inside the stenosphere; they are impelled inward (!), that is, in the direction of decreasing values of the parameter, r. In the VRM interpretation, the mystery disappears: The bodies are impelled, as always, in the direction from the concave to the convex side of the orbit. Inside the stenosphere, the force may be said to be directed "outward into innerspace."

Marek Abramowicz is largely responsible for bringing this important phenomenon to the attention of the scientific community.[10] He has made extensive studies of this effect, and has defined an "optical geometry" which is identical to the telemetric geometry of the VRM interpretation in the case of static fields (fields that are time independent, with no mixed space-time components in the expression for the invariant interval). Abramowicz has proved that inertial forces occur only when the motion of a body deviates from these light-ray geodesics. The fact that the telemetric geodesics correspond to the paths taken by light rays, follows from Fermat's principal that light rays follow the path of least time, and the fact that in the telemetric system, distance is measured by echo ranging.

## 13 Light Rays and Geodesics

Why light rays fail to follow proper geodesics is another puzzling question that arises in the standard interpretation of the theory based upon proper measurements. From the VRM point of view, the answer is evident: proper geodesics are defined as the shortest paths as measured by local observers using local measuring rods; these "shortest" paths in proper measure "cheat" by deviating inwardly so as to take advantage of the length dilation effect, which results in a smaller value for the distance measured.

Regarding the Shapiro effect, it is clear that the proper distance measured along the path of a light ray (a geodesic in the telemetric system, but not in proper measure) will be less than that measured by telemetric methods. It is also true that the time of flight is also underestimated, but to a lesser degree, since the time measurement is made at one end of the ray path. Hence, it is clear that in the usual interpretation, the speed of light must appear to be reduced in a gravitational field. Of course, in the telemetric system the time required for the radar signal to complete its journey is equal to the distance traveled divided by the constant speed of light as a matter of definition.

## 14 Newton's G, Einstein's Λ

An interesting issue arises when one considers the question as to whether the periods of all clocks are affected in the same way as are those of atomic clocks. For immediately, one thinks of a pendulum clock, the period of which is proportional to $\sqrt{(L/g)}$, where L is the length of the pendulum and g is the acceleration of gravity. Note that the "clock" being considered here includes the body producing the acceleration g, as well as the pendulum. Thus we may, for instance, consider how the rate of a pendulum clock operating on the Earth is influenced by the Earth-Sun distance.

Now g depends upon G, the Newtonian gravitational constant, and simple dimensional analysis shows that G has the "units" $L^3 T^{-2} M^{-1}$. Since $L^* \sim f^{-1}$, $T^* \sim f^{-1}$, and $M^* \sim f$, one must have $G^* \sim f^{-3} \cdot f^2 \cdot f^{-1} = f^{-2}$, which is properly written as $G^* = f^{-2} G$. From this one would conclude that a pendulum clock will run at the same rate as an atomic clock only if $G^*$ is strongly increased in a gravitational field. If, on the other hand, $G^*$ is actually a constant, it would mean that the result of a Cavendish-type experiment would depend upon position in a gravitational field, a result that would directly violate the strong principle of equivalence, which prohibits a free falling observer from detecting any gravitational effects. Any wise person will bet on the venerable principle of equivalence, but certainly, experiments should be carried out to verify that G, in local proper measure, is indeed constant.

In this connection, it is interesting to note that if Einstein's field equations are to retain their form under a conformal transformation (such as relates the proper and telemetric systems) with both G and Λ as actual constants, independent of gravitational potential, it is required that

$$\Lambda^* = f^{-2}\Lambda \text{ as well as } G^* = f^{-2} G$$

where Λ is the controversial cosmical constant.

## 15  A Heuristic Argument for Variable Rest Mass

This investigation began with the search for the cause of gravitational clock slowing.  It is also possible to make a direct approach by simply refusing to accept the dogma that gravitational forces are uniquely different from all other forces. In this regard Richards,[11] et al, assert

> ".the work of all forces acting on a body, *with the exception of the gravitational force*, equals the change in the total mechanical energy of the body." (emphasis theirs)

In addition to rejecting this assertion, the equivalence of mass and energy is invoked, and it is assumed that when a body is raised in a gravitational field, its mechanical energy is increased.  Further, it is assumed that this increase in energy will result in an increase in its rest mass in accord with Einstein's famous equation, $E=mc^2$.

It is easy (see Appendix B) to show from the integral of motion,
(viz., $\sqrt{(1-r_s/r)}/\sqrt{(1-v^2/c^2)}$ = const.) that the telemetric acceleration, $g^*$, of a body falling from rest at r, has the initial value

$$g^* = d^2R^*/dt^2 = GM/r^2$$

Then we can write

$$dm^*c^2 = m^* g^* dR^* = (GMm^*/r^2)dR^*,$$

but $$dR^* = dr / (1-r_s/r),$$

so $$dm^*/m^* = (GM/c^2)/r^2 \, dr / (1-r_s/r) = (1/2) \, r_s/r^2 \, dr / (1-r_s/r)$$
$$= d[\sqrt{(1-r_s/r)}] / [\sqrt{(1-r_s/r)}]:$$

whence $$m^* = m \sqrt{(1-r_s/r)}.$$

Thus, the direct approach yields a result consistent with that found by the original approach, namely, the systematic analysis of the cause of the gravitational slowing of clocks, which, unfortunately, is very often called the "gravitational time dilation effect."

## 16  The Kerr Solution

The VRM interpretation may be extended to fields that involve steady-state motion.  Thus, it provides insight into phenomena occurring in the Kerr solution (the rotating black hole); for example, the (unstable) photon orbits can be shown to lie on circles whose curvature matches that induced in the ray paths by the shear in the frame-dragging velocity of the vacuum.[12]

## 17  Extension to Cosmology

Looking back, it may be said that the foundation of the VRM interpretation, as applied to stationary fields, is the conservation of energy.  It was not until 1918 that the connection between conservation laws and the symmetries of a system was elucidated by the great algebraist, Emmy Noether.  The connection is simply stated:  For each symmetry, there is one conserved quantity.  In the case of the Schwarzschild solution, the time-independence of the field gives rise to the conservation of energy.  Similarly, the angular symmetry of the field results in the conservation of angular momentum.  It turns out that the product of the dimensionality of the coordinate of symmetry and that of the conserved quantity always has the dimension of action, $ML^2T^{-1}$.

In the case of the solutions thought to be descriptive of the expanding universe, typified by the Friedman solutions, the important symmetry is spatial homogeneity, the corresponding coordinate of which has the character of length. The conjugate variable of length is momentum. It is to be expected, then, that in a Friedman-type universe, momentum will be conserved. Solving the equation of motion for a freely moving test mass in the Friedman universe is easily done (see Appendix C), and the result may be written

$$A(t) \beta/\sqrt{(1-\beta^2)} = \text{constant}$$

Here $\beta = v/c$, is the relativity parameter, and $A(t)$ is the function describing the manner in which the universe expands with the time, t. Clearly as the universe expands, $\beta$, the relativity parameter of the test mass must decrease. However, since the momentum is given by $p = mc \beta/\sqrt{(1-\beta^2)}$, this would seem to be inconsistent with conservation of momentum: unless, that is, the rest mass of the particle were to increase in proportion to the function $A(t)$. Now it is not possible that the proper rest mass increases, since proper masses are defined in terms of prototype masses, such as that of the carbon 12 atom, or that of a certain Platinum-Iridium cylinder. On the strength of the analogy with the situation that pertains for the Schwarzschild field, it seems reasonable to take a speculative leap, and assume that the VRM rest mass, m*, of any body is related to its constant proper rest, m, by the relation,

$$m^* = A(t)/A_{pe}\, m\,,$$

where $A_{pe}$ is the value of A at some fixed reference time, say that of the present epoch. Multiplying the integral of motion by the constant $mc/A_{pe}$, we have

$$m^*c\beta/\sqrt{(1-\beta^2)} = p^* = \text{constant},$$

which says that VRM momentum is conserved.

## 18  Cosmic Length Contraction and the Hubble Red shift

Now, recalling that atomic dimensions are inversely proportional to rest mass, it is clear that in the VRM interpretation, the length of measuring rods will shrink as $A_{pe}/A(t)$! Thus, according to this VRM interpretation, the expansion of the universe is only an apparent phenomenon; the VRM "truth" is that we, along with our measuring rods, are contracting!
Of course, the increase of rest masses also means that the rate of every clock constructed of ordinary matter is continually increasing. In this interpretation, the red shift has nothing to do with the recession of distant galaxies: it is simply a consequence of the fact that the frequency of any given atomic transition was, in past eras, reduced as compared with the present values for the same transition.

## 19  A Cosmic Clock

It is easily shown that in the Friedman cosmology the wavelength of free radiation bears a constant ratio to the scale function, $A(t)$. Thus in the usual (proper) interpretation of the general theory, it is correct to say that the free radiation expands as the universe expands. Such is the case with the cosmic background radiation (CBR).
The implementation of a telemetric-type system requires a standard for timekeeping that is not influenced by the change of rest mass of ordinary matter. But momentum conservation, applied to the photons of the CBR, implies that their frequency must, in the VRM interpretation, be constant. Hence, for the immensely impractical purposes of cosmology, the CBR will serve in this role. Thus, one might take the peak of the CBR blackbody spectrum, approximately 300 GHz, as a frequency standard for a reference clock.

## 20  The Non-Expanding Universe

If one were able to perform echo ranging experiments with distant galaxies, it would turn out that the distance computed, based upon CBR time, would remain constant for galaxies having no peculiar motion. Thus, in the cosmical VRM interpretation, the universe has a fixed size.

As one looks back in time, the dimensions of all material bodies continually increase, and the rates of all material processes slow. Furthermore, as a consequence of the conservation of momentum, the relativity parameter, $\beta$, associated with any peculiar motion, however tiny in the present epoch, will continually increase, and will eventually go to one as $A(t)$ approaches zero. Thus, the behavior of matter as one looks back in time to the "Big Bang," may be said to mimic the behavior of matter falling inward toward the event horizon of a black hole. The analogy is not perfect, however, for it turns out that a detailed analysis of the behavior of the function, $A(t)$, shows that, for nearly any model for the equation of state of matter, the function $A$ vanishes at a finite epoch, even when time is measured using the CBR standard.

## 21  Not with a Bang

In the cosmical VRM interpretation, the universe had a beginning, but it was a very quiet affair. In the earliest epoch, matter of every type was all of a radiant form, each type a quantum field possessing only an infinitesimal rest mass, tending to zero. Something, perhaps a scalar Higgs field, initiated the processes that caused certain of these fields to develop mass, and so to differentiate into the various forms of matter that make up the universe as we now perceive it.

This process may have occurred very rapidly, especially in terms of proper time, without any need for the faster-than-light expansion of space itself that is a key feature of inflationary theories of the origin of the universe. It is also notable that the troubling space-time singularity of the Big Bang, inherent in the standard interpretation (at least for simple, highly symmetric models), does not appear at all in the VRM interpretation.

## 22  Conclusions: A Deeper Understanding

Summing up, it may be said that the Variable Rest Mass interpretation possesses several advantages over the usual interpretation.

With regard to stationary fields, it reveals a satisfying harmony with older Newtonian concepts. Thus, gravitational rest mass reduction may be said to be the literal embodiment of the concept of gravitational potential energy. In the same vein, the principle of energy conservation, employing the customary definition of energy, is retained. It may also be remarked that the VRM interpretation eliminates the rather unsettling need to ascribe a negative energy density to static gravitational fields.

It also eliminates the need to deal with the conceptual difficulties of the infinite mass density and, indeed, the singularity, which in the standard interpretation, seem to exist inside the event horizon of a black hole.

In addition, the VRM interpretation explains why light rays do not correspond to proper geodesics; it explains the Shapiro effect without requiring the embarrassing need to contemplate a reduction in the speed of light; and perhaps most notably, makes comprehensible the otherwise incredible Abramowicz effect.

More speculatively, as applied to cosmology, the VRM interpretation again re-connects with Newtonian concepts, conserving momentum in terms of the usual definition. It provides an explanation for the cosmic red shift that does not involve the rather mysterious concept of the expansion of space itself. Perhaps the best features of the interpretation are those that are not present, namely, the infinite densities and singularity of the big bang, and the faster-than-light expansion of space itself, that, in the standard interpretation, is an unavoidable feature of inflationary theories.

However, as regards the evolution of the universe, the *cause* of the increase of rest masses proposed in the present interpretation is no less a mystery as is the expansion of space in the big bang interpretation. Also, it must be admitted that that the VRM program appears to be limited to those situations, such as stationary fields, or cosmologies of the Robertson-Walker type, in which one can define a world time. Nevertheless, it seems clear that the dependence of rest mass upon the gravitational field is a universal and important phenomenon.

**Appendix A**

Let $c^*/c = f^{(c)}$, $h^*/h = f^{(h)}$, and $m_e^*/m_e = f^{(m)}$, where $f = \sqrt{(1 - 2GM/c^2 r)}$. It will be assumed that the Rydberg frequency, $R_\infty c \sim m_e e^4/h^3$ and the Bohr radius $a_0 \sim h^2/m_e e^2$ characterize instruments for the measurement of time and distance, respectively.

Gravitational time dilation requires that $(R_\infty c)^*/(R_\infty c) \sim f$. Assuming that the electronic charge is invariant, this implies $(m) - 3(h) = 1$.

Regarding the fact that local observers measure the speed of light to be constant, note that distance/time $\sim a_0 R_\infty c$. Hence $c_{meas} \sim a_0 R_\infty c$, and consequently $c_{meas} = c$ implies that $a_0 R_\infty \sim e^2/hc$, the fine structure constant, is invariant. Thus $(a_0 R_\infty)^*/(a_0 R_\infty) = (e^2/hc)^*/(e^2/hc) \sim f^0$
Again taking $e^*$ to be invariant, we have $(h) + (c) = 0$.

Summing up, the conditions that must be satisfied among the exponents (c), (h) and (m) are

$$(c) = -(h) = (1/3)[1 - (m)].$$

**Appendix B**

For a body falling from rest at $r_0$ in the Schwarzschild field, the integral of motion is
$\sqrt{(1-r_s/r)} / \sqrt{(1-v^2/c^2)} = \sqrt{(1-r_s/r_0)}$, whence, $v = c \sqrt{(r_s/r - r_s/r_0)} / \sqrt{(1-r_s/r_0)}$.
Now $v^* = v$, and $g^* = dv^*/dt$. Thus $g^* = dv/dt = -\frac{1}{2} c r_s/r^2 / [\sqrt{(r_s/r - r_s/r_0)} (1-r_s/r_0)] \, dr/dt$.
But $v = v^* = dR^*/dt = (1-r_s/r)^{-1} dr/dt$, whence
$dr/dt = (1-r_s/r) v = c (1-r_s/r) \sqrt{(r_s/r - r_s/r_0)} / \sqrt{(1-r_s/r_0)}$.
Inserting this into the expression for $g^*$ yields $g^* = -\frac{1}{2} c^2 r_s/r^2 (1-r_s/r) / (1-r_s/r_0)$,
and finally, $g^*(r_0) = -\frac{1}{2} c^2 r_s/r_0^2 = -GM/r_0^2$

**Appendix C**

The Friedman metric may be written (ref [9], p. 380), substituting a(t) in place of A(t)

$$ds^2 = c^2 d\tau^2 - a^2(\tau)[\, d\chi^2 + s^2(\chi)(\, d\theta^2 + \sin^2\theta \, d\phi^2)]$$

where the function $s(\chi)$ is $\sin(\chi)$, $\chi$, or $\sinh(\chi)$ depending on whether the Hubble constant is less than, equal to, or greater than $\sqrt{(8\pi/3 \cdot G\mu)}$. Here $\mu$ is the equivalent mass density of the total mass-energy density in the universe.

Introducing a new time variable, $\eta$, defined by $cd\tau = ad\eta$, the metric becomes

$$ds^2 = a^2(\eta)\,[\, d\eta^2 - d\chi^2 - s^2(\chi)(\, d\theta^2 + \sin^2(\theta) \, d\phi^2)]$$

Without loss of generality, one may consider motion in the $\chi$ direction, for which the geodesic equation is

$$d^2\chi/ds^2 + \Gamma^\chi_{\eta\eta}(d\eta/ds)^2 + 2\Gamma^\chi_{\eta\chi}(d\eta/ds)(d\chi/ds) + \Gamma^\chi_{\chi\chi}(d\chi/ds)^2 = 0$$

But $\Gamma^\chi_{\eta\eta} = 0$, $\Gamma^\chi_{\eta\chi} = a^{-1}\, da/d\eta$, and $\Gamma^\chi_{\chi\chi} = 0$. Thus the geodesic equation reduces to

$$d^2\chi/ds^2 + 2a^{-1}(da/d\eta)(d\eta/ds)(d\chi/ds) = 0$$

Integrating, using $(da/d\eta)(d\eta/ds) = da/ds$, produces

$$a^2 \, d\chi/ds = \text{constant.}$$

The proper velocity, $v$, is defined by $v = dl/d\tau$, where $dl = ad\chi$ for the motion considered, and $d\tau = a/c \, d\eta$. Thus $v/c = d\chi/d\eta$, and the integral may be written

$$a^2 \, d\chi/ds = a^2 \, (d\chi/d\eta)/(ds/d\eta) = a^2 \, (v/c)/(ds/d\eta) = \text{constant}$$

But directly from the line element, for the motion considered, one has

$$ds/d\eta = a\, \sqrt{[1-(d\chi/d\eta)^2]} = a\, \sqrt{[1- (v/c)^2]}$$

so that finally the integral of motion becomes

$$a(\eta)\,(v/c) / \sqrt{[1 - (v/c)^2]} = \text{constant}$$